\documentclass[11pt]{article}

\usepackage[preprint]{acl}
\usepackage{paralist}
\usepackage{times}
\usepackage{latexsym}

\usepackage[T1]{fontenc}
\usepackage{amssymb}
\usepackage[utf8]{inputenc}

\usepackage{microtype}

\usepackage{inconsolata}

\usepackage{graphicx}

\usepackage[most]{tcolorbox}
\usepackage{listings} 
\usepackage{xcolor}   
\usepackage{siunitx}
\usepackage{booktabs}
\usepackage{multirow}
\tcbuselibrary{breakable, skins} 
\tcbset{
    substep_clean/.style={
        enhanced,
        colframe=blue!15!white,      
        colbacktitle=blue!5!white,   
        coltitle=black,              
        colback=white,               
        fonttitle=\bfseries\small,   
        boxrule=0.5pt,               
        arc=2pt,                     
        titlerule=0.0pt,             
        left=6pt, right=6pt, top=4pt, bottom=4pt, 
        boxsep=2pt,
        before skip=6pt,             
    }
}

\usepackage{enumitem}
        
\title{Enhancing Local Life Service Recommendation with \\Agentic Reasoning in Large Language Model}

\author{
 \textbf{Shiteng Cao \textsuperscript{1,*}},
 \textbf{Xiaochong Lan\textsuperscript{2,*}},
 \textbf{Yuwei Du\textsuperscript{2}},
 \textbf{Jie Feng\textsuperscript{2}},
\\ \textbf{Yinxing Liu\textsuperscript{3}},
 \textbf{Xinlei Shi\textsuperscript{3}},
 \textbf{Yong Li\textsuperscript{2}},
\\ \textsuperscript{1} Shenzhen International Graduate School,Tsinghua University\\
 \textsuperscript{2} Department of Electronic Engineering, BNRist,
Tsinghua University\\
 \textsuperscript{3} Meituan\\
\textsuperscript{*} Equal contribution \\
\small{
   \textbf{Correspondence:} liyong07@tsinghua.edu.cn,fengj12ee@hotmail.com
 } }

\begin{document}

\maketitle

\begin{abstract}
Local life service recommendation is distinct from general recommendation scenarios due to its strong living need-driven nature. Fundamentally, accurately identifying a user's immediate living need and recommending the corresponding service are inextricably linked tasks. However, prior works typically treat them in isolation, failing to achieve a unified modeling of need prediction and service recommendation. In this paper, we propose a novel large language model based framework that jointly performs living need prediction and service recommendation. To address the challenge of noise in raw consumption data, we introduce a behavioral clustering approach that filters out accidental factors and selectively preserves typical patterns. This enables the model to learn a robust logical basis for need generation and spontaneously generalize to long-tail scenarios. To navigate the vast search space stemming from diverse needs, merchants, and complex mapping paths, we employ a curriculum learning strategy combined with reinforcement learning with verifiable rewards. This approach guides the model to sequentially learn the logic from need generation to category mapping and specific service selection. Extensive experiments demonstrate that our unified framework significantly enhances both living need prediction performance and recommendation accuracy, validating the effectiveness of jointly modeling living needs and user behaviors.
\end{abstract}




\section{Introduction}
Local life service platforms, such as Meituan\footnote{meituan.com}, facilitate urban residents in the discovery and acquisition of surrounding physical services, including dining, accommodation, domestic help, and entertainment~\cite{li2022automatically,ping2021user}. These platforms cover the core scenarios of daily urban life. Building upon this, the central task of local life service recommendation is to precisely recommend merchants that satisfy the user's potential living needs, taking user profiles, preferences, and spatiotemporal contexts as input. Such recommendations aim to fulfill real-world living needs and directly lead to offline physical consumption, thereby exerting a substantive influence on users' daily lives. Precise recommendations significantly reduce decision-making costs, helping individuals efficiently obtain high-quality experiences.

Distinct from general recommendation scenarios, local life services are characterized by a strong living need-driven nature.
Fundamentally, living need prediction and service recommendation have an inseparable correlation. Both must be rooted in the comprehensive modeling of users and contexts, as well as a deep understanding of human need generation and behavioral laws. This correlation encourages the joint modeling of these two tasks, with the expectation that they will mutually enhance one another. However, prior works have not achieved a unified modeling of need prediction and service recommendation. Existing industrial generative methods for local life service recommendation~\cite{han2025mtgr,zhou2025onerec,zhou2025onerecv2} do not explicitly model user living needs. 
Conversely, existing living need prediction methods~\cite{lan2023neon,lan2025open} explicitly model needs, yet the subsequent mapping between needs and service recommendation relies on hard rules or separately trained embedding models. This lack of joint modeling constrains the accuracy of the final recommendation. This points to a significant research gap: how to construct a unified framework that jointly performs living need prediction and service recommendation.

Large Language Models (LLMs) offer a new opportunity to address this gap~\cite{xi2023KAR,chen2024hllm,liu2025larm}. LLMs possess extensive world knowledge and commonsense reasoning capabilities, enabling a deep semantic understanding of human living needs. Furthermore, they can unify the prediction of needs and services within a single linguistic format. By sharing the same token space, LLMs can naturally model the joint distribution of living needs and services. Nevertheless, this process presents challenges:

Firstly, constructing high-quality training data that reflects the true laws of need-driven consumption is difficult. Raw records from platforms do not centrally reflect the logic of need generation. The reasons behind recorded consumption behaviors are manifold; beyond modelable laws of living needs, records contain noise from unpredictable factors such as platform promotions. Screening for data that reflects universal laws of need generation is essential for effective model learning but constitutes a significant challenge.

Besides, the joint modeling of living needs and specific services involves navigating an extremely vast search space. This complexity stems from the high diversity of fine-grained living needs, the massive scale of available merchants, and the numerous logical paths that explain how a specific service satisfies a need. Consequently, learning to accurately map abstract need generation to specific service interaction within this expansive space presents a significant difficulty.

To resolve these challenges, we propose a novel joint modeling method. To address the first challenge, we design a clustering method based on user behavioral features. This approach filters out accidental noise and selectively preserves the behaviors of typical users, allowing the model to focus on general behavioral laws. Unlike traditional machine learning methods that require explicit memorization of long-tail data, our approach enables the LLM to learn a compact yet robust logical basis from high-quality patterns. Relying on the model's inherent reasoning capabilities, it can then effectively generalize this learned logic to long-tail scenarios. To address the second challenge, we adopt a curriculum learning approach during training. We constrain the search space by allowing the model to learn in steps. Specifically, the training phase is designed to let the model sequentially learn the laws of need generation, the mapping from need to category, and finally, specific user behavior generation. Additionally, we employ Reinforcement Learning with Verifiable Rewards (RLVR) rather than simple imitation learning to enhance the model's active reasoning ability. In the inference phase, our model adopts an agentic reasoning method, simulating the human cognitive process from living need to specific consumption behavior, thinking step-by-step to solve the recommendation problem.

Our contributions are as follows:
\begin{inparaenum}[(1)]
    \item We are the first to propose the unified problem of user living need and behavior prediction in local life service recommendation.
    \item We develop a novel methodology that integrates clustering-based sampling to capture robust behavioral laws and employs curriculum learning with RLVR to effectively constrain the vast search space. This enables the model to master the agentic reasoning process from abstract living needs to specific service interactions.
    \item We conduct extensive experiments to verify our framework, demonstrating its superior performance over existing baselines and validating the effectiveness of each component.
\end{inparaenum}

\section{Problem Formulation}

\begin{figure*}[t]
\centering
\includegraphics[width=0.98 \linewidth]{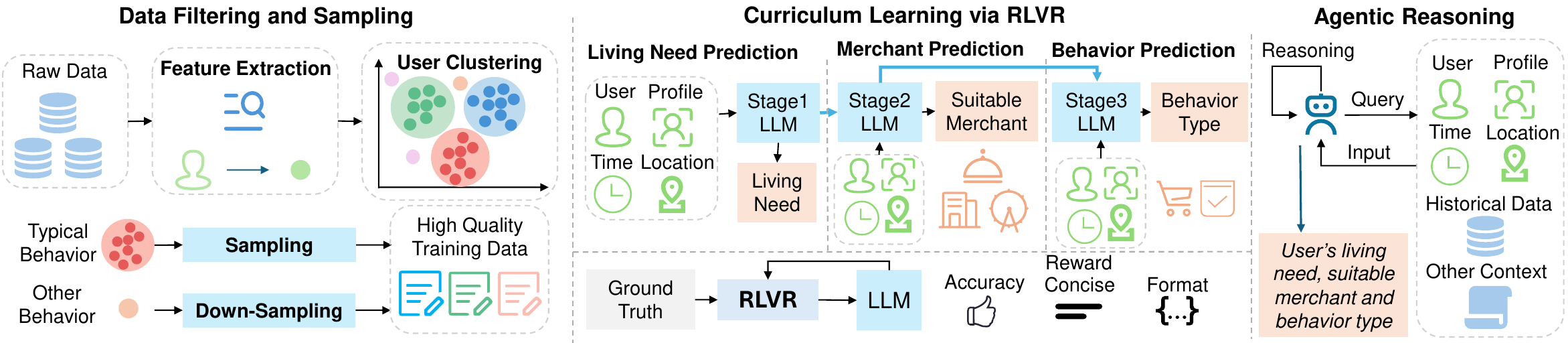}
\caption{The overall architecture of HiAgentRec, which integrates a noise-robust data pipeline with an RLVR-optimized curriculum learning strategy for hierarchical reasoning.}
\label{fig:overall_framework}
\end{figure*}
We formulate local life service recommendation not merely as a sequential matching task, but as a hierarchical agentic decision-making process driven by explicit living needs. We define the hierarchical decision space through two intermediate granularities. The highest level consists of living needs $\mathcal{I}$, which represent high-level latent intents such as \textit{Business Travel}. Given that such needs are typically implicit in raw click logs, we construct the ground truth set $\mathcal{I}$ by leveraging LLMs to extract explicit need descriptions from user post-transaction reviews. Bridging these abstract needs and specific merchants, we define Semantic Categories $\mathcal{C}$ as the set of coarse-grained service types, such as \textit{Hotel}, which facilitate the mapping from intent to concrete consumption behaviors.

For each user $u \in \mathcal{U}$, we construct a comprehensive context view comprising three integral components. First, the user profile $\mathcal{P}_u$ encompasses static feature sets. Second, the historical sequence $S_u$ records the chronological history of past interactions, denoted as $S_u = \{y_1, y_2, \dots, y_n\}$, where each interaction $y_k = (i_k, c_k, b_k)$ is explicitly modeled as a tuple containing the living need, semantic category, and specific behavior. Finally, the spatio-temporal context $X_{n+1}$ captures the environmental constraints for the target prediction, specifically the timestamp $t_{n+1}$ and location $\ell_{n+1}$. Formally, given the user profile $\mathcal{P}_u$, historical sequence $S_u$, and the target context $X_{n+1}$, the primary objective of HiAgentRec is to model the joint probability of the user's hierarchical decision path.

\subsection{Related Work}
\subsubsection{LLM for Recommendation}

With the remarkable general capabilities demonstrated by LLMs such as ChatGPT~\cite{openai2023gpt}, DeepSeek~\cite{guo2025deepseek} and Qwen~\cite{qwen2024}, researchers have increasingly focused on how to harness LLMs for recommender systems. Existing approaches can be broadly categorized into two paradigms.

The first treats the LLM as an enhancer. For instance, KAR~\cite{xi2023KAR} leverages LLMs to generate textual descriptions of users or items and enhances the representation power of traditional recommendation models via embedding augmentation strategies. The second paradigm frames recommendation as a language processing task. Works such as P5~\cite{geng2023P5}, ONCE~\cite{liu2023once}, UniSRec~\cite{hou2022unisrec}, and M6Rec~\cite{cui2022m6rec} reformulate recommendation objectives into natural language prompts or generation tasks, enabling end-to-end LLM-based recommendation. However, a significant gap exists between the pretraining objective of LLMs and the optimization goals in recommendation. This mismatch often leads to challenges such as popularity bias and limited fine-grained controllability when LLMs are applied directly to recommendation tasks.~\cite{hua2023tutorial}

To bridge this gap, fine-tuning has become a widely adopted strategy, exemplified by representative works such as BIGRec~\cite{bao2023bigrec}, Softmax-DPO~\cite{chen2024softmax-dpo}, and RosePO~\cite{liao2024rosepo}. More recently, researchers have explored alignment techniques better suited for complex reasoning and multi-objective optimization. GRPO~\cite{Shao2024GRPO} marks a paradigm shift toward leveraging complex, multi-level feedback signals in LLM alignment. OneRec-Think~\cite{liu2025onerecthink} proposes a three-stage Reason–Reflect–Revise generation framework that explicitly guides the LLM to simulate the deep logic underlying user living need, item attributes, and contextual interactions. Rec-R1~\cite{lin2025recr1} establishes a general closed-loop optimization framework that directly utilizes feedback from a fixed recommender to refine LLM generations, bypassing the need for costly data distillation. RecLLM-R1~\cite{xie2025RecLLM-R1} combines GRPO with chain-of-thought reasoning and employs a flexible multi-objective reward function to jointly optimize accuracy, diversity, and novelty. STREAM-Rec~\cite{zhang2025slowthinkingsequentialrecommendation} integrates GRPO with explicit reasoning paths, transforming the recommender from a black-box matcher into an interpretable, multi-step decision process. GREAM~\cite{hong2025gream} enhances GRPO with structured semantic indexing and sparse reward augmentation to mitigate feedback sparsity and semantic drift in recommendation. Finally, RankGRPO~\cite{zhu2025rankgrpo} adapts the GRPO framework to naturally align its reward mechanism with the rank-style output inherent to recommendation tasks. However, existing approaches largely neglect the explicit modeling of user needs, which serve as the fundamental driving force in local life service scenarios. To bridge this gap, our work introduces a joint modeling paradigm that simultaneously optimizes the prediction of underlying living needs and concrete interactions.

\subsubsection{Local Life Service Recommendation}
Local life service recommendation serves as a crucial bridge connecting users to offline physical services and attracts significant attention from both academia and industry. Early research primarily relies on collaborative filtering and matrix factorization techniques~\cite{bao2015recommendations,lian2014geomf}. Subsequent methods employ Recurrent Neural Networks (RNNs) and Transformer architectures to capture user sequential behavioral patterns, and they further integrate spatio-temporal contexts specific to local life scenarios, such as geographical location and time slots, to enhance recommendation accuracy~\cite{feng2018deepmove,ni2020spatio}. More recently, studies adopt generative recommendation paradigms. By encoding items into semantic identifiers or generating target items directly, these paradigms bridge the semantic gap between user context and item representations to achieve greater precision~\cite{zhou2025onerec,zhou2025onerecv2,han2025mtgr}.

However, these mainstream methods largely focus on mining implicit preferences from historical interaction records. Essentially, they perform behavior-to-behavior fitting and overlook the core motivation driving local life consumption, specifically the rigid and explicit living need of users. To address this limitation, another line of research introduces living need recognition and living need prediction. These approaches utilize techniques such as disentangled representation learning or graph neural networks to infer latent user living needs~\cite{ping2021user,li2022automatically}, and recent works further attempt to explicitly model user living need states~\cite{lan2023neon,lan2025open}. Although these attempts improve system interpretability to some extent, they typically treat living need prediction and service matching as disjoint stages. The subsequent mapping process often relies on hard-coded rules or independently trained vector space retrieval, which prevents the effective translation of living need understanding into precise consumption decisions. Furthermore, while the application of large language models in recommender systems rises, typically for feature augmentation or ranking optimization, few studies achieve end-to-end joint modeling from abstract living needs to specific service consumption within a unified generative framework. Targeting this gap, our work proposes an LLM-based joint modeling approach. By simulating the human-like agentic reasoning process, our method achieves mutual enhancement between living need prediction and service recommendation.

\section{Methodology}
\subsection{Overall Framework}

In this section, we present \textbf{HiAgentRec} (\textbf{Hi}erarchi-cal \textbf{Agent}ic \textbf{Recommendation}), an RL-enhanced agentic framework designed to jointly perform living need prediction and service recommendation. As illustrated in Figure~\ref{fig:overall_framework}, HiAgentRec orchestrates a coarse-to-fine reasoning process. Our methodology comprises three integral components:

Addressing the high noise and sparsity inherent in user behavioral data, we introduce a \textit{Data Filtering and Sampling Pipeline} (Section~\ref{sec:data_pipeline}). 
To navigate the vast solution space, we employ a Curriculum Learning Strategy optimized by RLVR (Section~\ref{sec:curriculum_learning}).
To simulate the human cognitive process from living need to specific consumption behavior, we reframe the recommendation task as a multi-step decision-making process via our \textit{Agentic Reasoning Pipeline} (Section~\ref{sec:agentic_reasoning}). 

\subsection{Noise-Robust Data Construction}
\label{sec:data_pipeline}
To mitigate the impact of stochastic noise and data sparsity, we propose a three-stage curation pipeline that filters accidental interactions while preserving representative long-tail patterns. We implement a clustering-based approach to disentangle unstructured noise from valid behavioral signals. By partitioning the user feature space $\{\mathbf{x}_u\}$ into $K$ clusters using MiniBatchKMeans~\cite{sculley2010minibatchkmeans}, we calculate a normalized typicality score for each user:
\begin{equation}
z_u = \frac{\| \mathbf{x}_u - \boldsymbol{\mu}_k \|_2}{\sigma_k}
\label{eq:anomaly_score},
\end{equation}
where $\boldsymbol{\mu}_k$ and $\sigma_k$ denote the centroid and intra-cluster standard deviation of cluster $C_k$. 
Users exceeding a deviation threshold are flagged as point-level outliers, while clusters falling below a minimum support threshold $\tau_{\text{min}}$ are pruned as non-representative edge cases.

To prevent the information loss in uniform sampling, we evaluate each cluster $C_k$ based on its scale $S_k$, cohesion $Q_k$ (inlier ratio), and semantic dominance. We discard low-quality clusters ($S_k < \tau_{\text{size}}$ and $Q_k < \tau_{\text{quality}}$) to eliminate noise. Conversely, we assign an enhanced sampling rate $r_{\text{high}}$ to long-tail clusters (small $S_k$ but high $Q_k$) to preserve niche behavioral patterns. The final dataset $\mathcal{D}_{\text{final}}$ is adaptively balanced to maintain structural cohesion and semantic diversity.

\subsection{Curriculum Learning via RLVR}
\label{sec:curriculum_learning}

To effectively navigate the search space and bridge the semantic gap between abstract user needs and specific items, we design a curriculum learning strategy optimized by RLVR. Instead of directly optimizing the complex end-to-end mapping, we decompose the generative process into a sequence of hierarchical subtasks. The training protocol sequentially activates three optimization phases, progressively expanding the generative scope from intents to specific decision paths.

The curriculum begins with the living need alignment phase, where the model focuses solely on the high-level semantic space. In this foundational stage, the objective is to maximize the likelihood of the correct living need description. This step ensures the model grounds its reasoning in user intent without the interference of massive item candidates. Building upon this semantic foundation, the second phase, category-constrained reasoning, extends the prediction target to include service categories. The model learns to predict the business taxonomy conditioned on both the context and the generated need, optimizing the joint probability. This intermediate step bridges the gap between abstract intent and standardized platform supply. The process culminates in the full-path agentic decision phase. In this final stage, the model navigates the complete search space to identify the specific item, optimizing the full reasoning chain.

Throughout this curriculum, we employ GRPO~\cite{guo2025deepseek} as the optimization backbone. We initialize each new phase with the model checkpoint from the preceding stage, facilitating seamless knowledge transfer. This staged policy optimization allows the model to master the complex reasoning logic step-by-step, mitigating the issues of reward sparsity in direct end-to-end training.

We design a reward function \(\mathcal{R}\) to evaluate model outputs across a set of tasks For any task, the total reward is a weighted sum of matching accuracy, format validity, and response length. To ensure structural consistency and conciseness, we apply generic penalties across all tasks.
The Format Reward \(\mathcal{R}_{\text{fmt}}\) encourages valid JSON generation:
The Length Reward \(\mathcal{R}_{\text{len}}\) prevents verbose outputs. It is activated only when the prediction is correct (i.e., \(\mathcal{R}_{\text{match}} > 0\)) and decays over training steps to prioritize accuracy in later stages:
\begin{equation}
\begin{split}
    \mathcal{R}_{\text{len}} ={} & \mathbb{I}(\text{correct}) \cdot \alpha e^{-\frac{\text{step}}{T}} \\
    & \cdot \min\left(1, \max\left(0, 1 - \frac{\ell - \ell_{\min}}{\ell_{\max} - \ell_{\min}}\right)\right).
\end{split}
\end{equation}
Here, \(\mathbb{I}(\cdot)\) is the indicator function. This term creates a "safe zone" where \(\ell \leq \ell_{\min}\) yields maximum reward, penalizing only unnecessary tokens beyond \(\ell_{\min}\).

For category prediction, to account for linguistic ambiguity, we calculate the cosine similarity between the embeddings of the prediction and the ground truth. If the closest candidate in the taxonomy matches the label, a maximum reward of 1.0 is given; otherwise, the similarity score itself serves as a partial reward to guide the policy.

\subsection{Agentic Reasoning}
\label{sec:agentic_reasoning}
To transcend the limitations of black-box direct prediction, we propose a \textit{Hierarchical Agentic Reasoning Pipeline} that reframes next item prediction as an autonomous decision-making process. As illustrated in the framework, the LLM functions as an agent that actively orchestrates specific function tools to narrow down the solution space.

The process is governed by a System Role that defines the agent's objective: to formulate precise recommendations by analyzing user data and context. The pipeline enforces a structured interaction protocol. Specifically, the prompt guides the agent through a three-stage reasoning process: (1) Living Need Inference, which combines user profiles with spatiotemporal context to determine intent; (2) Category Mapping, which aligns identified needs with semantic domains (e.g., Food, Entertainment); and (3) Behavior Ranking, which selects the optimal specific action based on matching scores. To facilitate automated parsing, all agent outputs are strictly constrained to a JSON schema. Due to space limitations, the complete prompt specification and template are detailed in Appendix~\ref{sec:prompt_spec}.

\section{Experiments}
To evaluate our framework, we conduct extensive experiments focusing on seven research questions. We first assess the overall recommendation performance (RQ1) and the contribution of individual components (RQ2). We then investigate the model's robustness, specifically its improvement for cold-start users (RQ3). Furthermore, we provide case study regarding interpretability (RQ4), the preservation of general capabilities during specialization (RQ5), the mitigation of policy collapse (RQ6), and the impact of semantic embedding models (RQ7). Due to space constraints, the detailed results for RQ3–RQ7 are provided in the Appendix.

\subsection{Experimental Setup}
\begin{table*}[t]
\centering
\caption{Performance comparison on Shanghai (In-domain) and Beijing (Cross-City Generalization) settings. The best results are \textbf{bold}, and the second-best are \underline{underlined}.}
\label{tab:main_performance}
\resizebox{\textwidth}{!}{%
\begin{tabular}{l ccccc ccccc}
\toprule
\multirow{2}{*}{\textbf{Model}} & 
\multicolumn{5}{c}{\textbf{Shanghai (In-domain)}} & 
\multicolumn{5}{c}{\textbf{Beijing (Cross-City)}} \\
\cmidrule(lr){2-6} \cmidrule(lr){7-11}
 & HR@1 & HR@3 & HR@5 & NDCG@3 & NDCG@5 
 & HR@1 & HR@3 & HR@5 & NDCG@3 & NDCG@5 \\
\midrule
GRU4Rec & 0.3873 & 0.5182 & 0.5941 & 0.2982 & 0.2696 & 0.2840 & 0.4150 & 0.4560 & 0.3620 & 0.3810 \\
SASRec & 0.4387 & 0.6213 & 0.6984 & 0.5455 & 0.5773 & 0.3015 & 0.4381 & 0.4721 & 0.3842 & 0.3982 \\
BERT4Rec & 0.4685 & 0.6185 & 0.6973 & 0.5558 & 0.5881 & 0.3202 & 0.4504 & 0.4733 & 0.3963 & 0.4057 \\
LLM (Base) & 0.2160 & 0.2190 & 0.2240 & 0.2176 & 0.2209 & 0.1972 & 0.2120 & 0.2290 & 0.1922 & 0.2059 \\
LLM+SFT & 0.3879 & 0.5420 & 0.6650 & 0.4920 & 0.5280 & 0.2583 & \underline{0.4650} & \underline{0.5410} & 0.3820 & 0.4100 \\
KAR & \underline{0.4850} & \underline{0.6410} & \underline{0.7150} & \underline{0.5720} & \underline{0.6010} & \underline{0.3350} & 0.4620 & 0.4890 & \underline{0.4100} & \underline{0.4250} \\
\midrule
\textbf{HiAgentRec} & \textbf{0.6098} & \textbf{0.7420} & \textbf{0.8010} & \textbf{0.6620} & \textbf{0.6950} & \textbf{0.6210} & \textbf{0.7580} & \textbf{0.8120} & \textbf{0.6750} & \textbf{0.7080} \\
\bottomrule
\end{tabular}
}
\end{table*}

\subsubsection{Dataset} 
To evaluate the effectiveness of HiAgentRec, we utilize two large-scale real-world datasets collected from Meituan, a leading local life service platform in China. The datasets specifically focus on user behaviors within Shanghai (used for in-domain evaluation) and Beijing (used for cross-city generalization), capturing the dense and complex spatiotemporal interaction patterns of metropolitan areas.
The raw data comprises user logs rich in contextual information, encompassing timestamps, GPS locations, categories, and explicit user living needs (e.g., dining, tourism). 
The detailed statistics of both datasets are summarized in Appendix~\ref{details of dataset}.

\subsubsection{Comparison methods}
To evaluate the effectiveness of HiAgentRec, we compare it against a set of baselines, ranging from traditional deep sequential models to recent LLM-based approaches. To ensure a fair comparison, all models were trained and evaluated on the same dataset. Due to space limitations,, detailed descriptions of these baseline models are provided in Appendix~\ref{subsec:baselines}.

\subsection{Overall Performance(RQ1)}

We measure the performance of HiAgentRec and baselines across two settings: in-domain (Shanghai), cross-city generalization (Beijing), as shown in Table~\ref{tab:main_performance}. 
HiAgentRec achieves superior performance across all metrics in both in-domain and cross-city settings.
In the Shanghai dataset, our model outperforms the strongest baseline  by a significant margin, improving 25.73\% in HR@1 and 15.64\% in NDCG@5. 
In the zero-shot Beijing setting, HiAgentRec demonstrates remarkable robustness, achieving a 85.37\% and 66.59\% gain in HR@1 and NDCG@5 respectively compared to the best-performing baselines. This stands in contrast to traditional methods, which suffer from  performance degradation when transferring between cities.
The reasoning framework mitigates overfitting to source-specific statistical correlations.
Traditional sequential models rely on local transition patterns inherent to Shanghai's urban layout, leading to failure when spatiotemporal contexts shift. In contrast, by modeling the invariant causal chain, HiAgentRec captures universal living logic, thereby achieving robust out-of-distribution generalization.

\subsection{Ablation Study(RQ2)}
To systematically verify the effectiveness of each component in HiAgentRec, we conduct ablation studies focusing on three key aspects: the hierarchical agentic reasoning structure, the data curation pipeline, and the choice of semantic reward model. Furthermore, we provide a detailed analysis of the curriculum learning dynamics to validate our multi-stage training strategy.

As summarized in Table~\ref{tab:ablation_study}, we compare the full HiAgentRec model against three variants. 
The variant \textit{w/o Agentic Reasoning}, which attempts to directly map context to categories without intent inference, suffers the most significant performance drop. 
This sharp decline confirms that decomposing the complex recommendation task into a chain is crucial. 
Removing the data filtering pipeline (\textit{w/o Data Selection}) leads to a noticeable degradation in recommendation quality, evidenced by a 7.95\% decrease in NDCG@5 compared to the full model. 
This indicates that raw behavioral logs contain substantial noise. Our filtering purifies the training signal, allowing GRPO to optimize policies based on high-quality, representative user patterns.
Eliminating the semantic modeling component  results in a considerable performance loss. This underscores the necessity of a semantic bridge between the model's generated text and specific item.

\begin{table}[t]
\centering
\caption{Ablation study on Agent Reasoning, Data Selection, and Semantic Model components. We compare the full HiAgentRec against variants without the reasoning chain, without the data filtering pipeline, and with a generic BERT encoder replacing the BGE model.}
\label{tab:ablation_study}
\setlength{\tabcolsep}{2.5pt} 
\resizebox{\columnwidth}{!}{
\begin{tabular}{l ccccc}
\toprule
\textbf{Variant} & \textbf{HR@1} & \textbf{HR@3} & \textbf{HR@5} & \textbf{NDCG@3} & \textbf{NDCG@5} \\
\midrule
\textit{w/o Agentic Reas.}  & 0.4828 & 0.5172 & 0.5517 & 0.5045 & 0.5149\\
\textit{w/o Data Selection}      & 0.5980 & 0.7220 & 0.7800 & 0.6116 & 0.6438 \\
\textit{w/o Semantic Model}        & 0.5209 & 0.6383 & 0.7475 & 0.6283 & 0.6339 \\
\midrule
\textbf{HiAgentRec (Full)}      & \textbf{0.6098} & \textbf{0.7420} & \textbf{0.8010} & \textbf{0.6620} & \textbf{0.6950} \\
\bottomrule
\end{tabular}

}
\end{table}

We further investigate the training dynamics to understand how Curriculum Learning facilitates model convergence.
The model trained in the preceding phase serves as a superior initialization checkpoint for subsequent tasks compared to the raw LLM. We observe that the initial accuracy for category prediction improves from 0.180 to 0.184. 
This empirically suggests that the preliminary task of inferring user living need equips the model with a common sense understanding of the domain, thereby laying a robust foundation for fine-grained category and behavior predictions.

Table~\ref{tab:curriculum_learning} illustrates the evolution of metrics during the Task 2 (Category) training phase. A compelling observation is the backward positive transfer: although Phase 2 focuses on optimizing Category output, the accuracy of the upstream \textit{Living need} task also improves steadily, rising from 51.36\% to 61.55\%. 
This indicates that the reasoning chain is not merely a one-way dependency; the feedback from fine-grained category matching reinforces the model's understanding of user living needs. 

\begin{table}[htbp]
\centering
\caption{Evolution of Living need Accuracy, Category Hit Rate (HR), and Ranking Quality (NDCG) during Task 2 training.}
\vspace{-0.2cm}
\label{tab:curriculum_learning}
\setlength{\tabcolsep}{2.5pt} 
\resizebox{\columnwidth}{!}{%
\begin{tabular}{l c ccc cc}
\toprule
\textbf{Stage} & \textbf{Need Acc.} & \textbf{HR@1} & \textbf{HR@3} & \textbf{HR@5} & \textbf{NDCG@3} & \textbf{NDCG@5} \\
\midrule
After Task 1 & 0.5136 & 0.1843 & 0.2512 & 0.2890 & 0.1954 & 0.2013 \\
20\% step    & 0.5291 & 0.3992 & 0.5215 & 0.5840 & 0.4125 & 0.4253 \\
40\% step    & 0.5500 & 0.5105 & 0.6420 & 0.7054 & 0.5350 & 0.5512 \\
60\% step    & 0.6318 & 0.5315 & 0.6750 & 0.7380 & 0.5827 & 0.6129 \\
80\% step    & 0.6227 & 0.5763 & 0.7180 & 0.8052 & 0.6350 & 0.6684 \\
After Task 2 & 0.6155 & 0.6028 & 0.7350 & 0.7945 & 0.6419 & 0.6874 \\
\bottomrule
\end{tabular}
}
\end{table}

\subsection{Case Study(RQ4)}
\begin{figure}[htbp]
  \centering
  \includegraphics[width=1.0\linewidth]{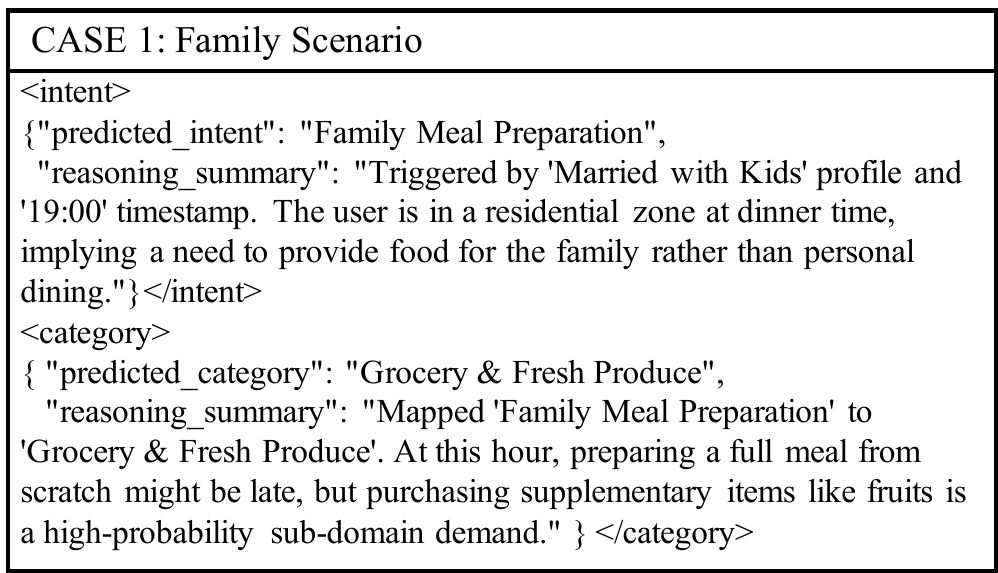}
\vspace{-0.6cm}
  \caption{Case Study 1: Cold-Start scenario where the agent infers based on the user's profile and time context.}
  \label{fig:case1}
  \vspace{-0.2cm}
\end{figure}

\begin{figure}[htbp]
  \centering
  \includegraphics[width=1.0\linewidth]{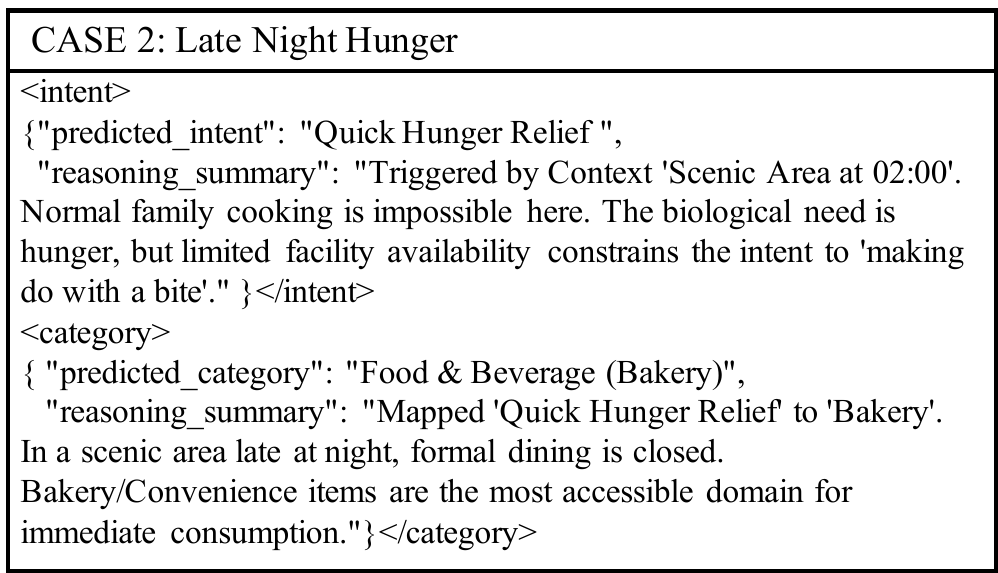}
  \vspace{-0.6cm}
  \caption{Case Study 2: Late-night hunger scenario showing how the agent handles cross-context reasoning.}
  \label{fig:case2}
  \vspace{-0.2cm}
\end{figure}

\begin{figure}[htbp]
  \centering
  \includegraphics[width=1.0\linewidth]{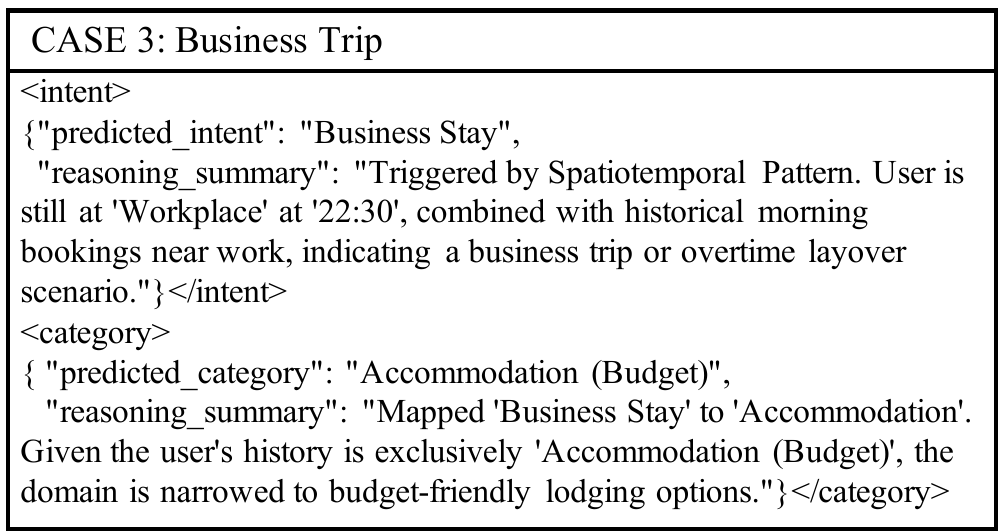}
  \vspace{-0.6cm}
  \caption{Case Study 3: Business trip scenario demonstrating the agent's ability to identify patterns from history.}
  \label{fig:case3}
\vspace{-0.2cm}
\end{figure}

To illustrate the reasoning capabilities of HiAgentRec under complex scenarios, we present three representative cases in Figures~\ref{fig:case1}, \ref{fig:case2}, and \ref{fig:case3}. These cases highlight HiAgentRec's ability to utilize world knowledge and logical deduction to handle cold-start and context-shifting scenarios effectively.

Case 1 (Cold-Start \& Zero-Shot): This user has no historical behavior. The model relies entirely on the profile (Married, Kids) and context (19:00, Residential Area). HiAgentRec  infers that 7 PM is a typical dinner preparation time for a family. It reasons that while main ingredients might be ready, Fruit serves as a healthy, high-probability supplement for children after dinner, accurately predicting the living need of Family Care.
    
Case 2 (Cross-Context \& Habitual): This user is in a generic Scenic Area at 2:00 AM, a scenario vastly different from their usual history (Home/Workplace). Traditional models might fail due to the location shift or recommend generic tourist services. HiAgentRec combines the context (late night + limited options) with the user's strong historical preference for Bread \& Cakes (Snack). It reasons that the user needs quick hunger relief  rather than a formal meal, correctly recommending a bakery item.
    
Case 3 (Specific Living need Recognition):This user appears at their workplace late at night (22:30). By analyzing the history of booking Economy Hotels near the workplace in the mornings, the model identifies a pattern of business travel rather than leisure. Consequently, it predicts an Accommodation living need for a budget-friendly hotel, aligning with the user's history.

\section{Deployment and Application}
To address the diverse and complex service requirements in Meituan’s local life ecosystem, we deployed the model within Meituan’s production ecosystem across two distinct high-traffic scenarios. Considering that user living needs typically exhibit spatio-temporal stability, where motivations remain relatively consistent within specific time windows and geographic areas, we implemented a near-line deployment strategy. Specifically, HiAgentRec triggers demand inference only when a shift in a user's spatio-temporal context is detected. These predictions are then integrated as a supplementary recall channel for the online recommendation pipeline.

The Huisheng (Smart Savings) prioritizes promotional value and cross-category discovery, serving as a critical traffic aggregation hub within Meituan’s system. By injecting intent-aware categories generated by HiAgentRec, we aimed to surface items that traditional collaborative filtering might overlook.
During a rigorous online A/B test from January 14, 2026, to February 3, 2026, involving $20\%$ of total production traffic, our proposed method achieved substantial performance gains over the highly optimized production baseline. Specifically, we observed a $13.12\%$ increase in CTCVR, a $4.37\%$ lift in CTR, an $8.32\%$ improvement in CVR, and a $10.11\%$ growth in GTV. These results underscore the model’s capacity to understand user needs, effectively scaling Meituan’s transaction volume.

The second deployment focuses on the landing page, where the primary objective is to maintain user engagement and facilitate conversion following an initial click. During a rigorous online A/B test from January 16, 2026, to February 5, 2026, involving $20\%$ of total production traffic, the deployment on the landing page same as Huisheng module yielded an overall CTCVR increase of $1.48\%$, an overall CVR increase of 1.46\% and a UV New User Conversion Rate (UV-NCR) growth of $2.79\%$. A granular analysis across vertical business sectors reveals the model's exceptional cross-domain adaptability. UV-NCR increased by $4.04\%$ in Leisure and Entertainment, $6.65\%$ in Beauty and Haircare, $2.84\%$ in Meituan Instashopping, $4.20\%$ in In-store Dining, and $2.74\%$ in Pharmaceutical services. These consistent improvements across diverse domains underscore the versatility and precision of the agent’s reasoning in capturing multifaceted user interests.

Beyond overall performance gains, HiAgentRec demonstrates efficacy in cold-start recommendation. By leveraging the semantic reasoning of LLMs, we observed a substantial lift in conversion metrics for new users and items. Specifically, the deployment resulted in a $1.45\%$ increase in Payment UV (Unique Visitors) and a $1.52\%$ rise in Payment PV (Page Views). For the new user segment, the New User Payment UV grew by $3.07\%$, while New User Payment PV increased by $2.90\%$. 

Overall, these experimental results demonstrate the practical value of our proposed method in large-scale industrial applications.

\section{Conclusion}
This paper proposes HiAgentRec, a unified framework for local life service recommendation that integrates living need prediction with agentic reasoning. By modeling the causal chain, the framework addresses the limitations of disjoint modeling in existing approaches. We introduce a clustering-based data filtering pipeline to mitigate noise and employ a curriculum learning strategy with RLVR to navigate complex search spaces. Extensive experiments on real-world datasets demonstrate that HiAgentRec consistently outperforms baselines. Notably, the model exhibits strong robustness in cross-city generalization and cold-start scenarios, validating the efficacy of the hierarchical reasoning structure. Future work will focus on incorporating multimodal information and exploring multi-objective optimization to further enhance recommendation performance.
\section{Limitations}
Despite the superior performance demonstrated by HiAgentRec, several limitations warrant discussion and provide directions for future work.

First, due to the computational overhead, our current deployment is restricted to near-line scenarios rather than real-time online inference. Future work will focus on model acceleration techniques, including knowledge distillation to smaller specialized models and structured pruning, aiming to achieve online inference capabilities.

Second, the framework currently relies exclusively on textual modalities, overlooking the rich multimodal information inherent in local life services. User reviews often contain images and merchants provide visual menus and storefront photos. Incorporating such multimodal signals could enhance the model's understanding of user living needs, particularly in scenarios where text alone is insufficient.
Addressing these limitations through continued research will further bridge the gap between innovation and deployment.

\bibliography{references}

\appendix
\newpage
\section{Appendix}

\subsection{Implementation Details}
We train the model for 2 epochs using a global batch size of 32, further divided into PPO mini-batches of size 128 and micro-batches of size 2 for gradient accumulation and memory efficiency.

Our base language model is Qwen2.5-3B-Instruct~\cite{qwen2025qwen25technicalreport}, which we fine-tune using the Volcano Engine Reinforcement Learning for LLMs~\cite{sheng2024verl}. The methodology is model-agnostic and can be readily adapted to other instruction-tuned language models.
To enable memory-efficient training, we employ Fully Sharded Data Parallel (FSDP). Gradient checkpointing and optimizer offloading are enabled to further reduce GPU memory consumption. The actor model is optimized using AdamW with a learning rate of $1 \times 10^{-6}$.

During policy rollouts, we generate responses using vLLM~\cite{kwon2023vllm} with the following sampling configuration: temperature $= 0.6$, top-$p = 0.95$, and $n = 16$ responses per prompt to ensure diverse reward estimation. For semantic matching in the reward model, we utilize the BGE-Large-Zh~\cite{bge_embedding} embedding model to compute semantic similarity between generated outputs and reference texts.
\subsection{Details of Baseline Models} 
\label{subsec:baselines}
To comprehensively evaluate the performance of HiAgentRec, we compare it against a diverse set of representative baselines. These methods cover three distinct paradigms: traditional sequential recommendation, direct LLM-based approaches, and LLM-enhanced frameworks.
\paragraph{Traditional Sequential Recommendation}
These models rely on ID-based embeddings and deep neural networks to capture sequential patterns.
\begin{itemize}[leftmargin=*]
    \item \textbf{GRU4Rec}~\cite{hidasi2016gru4rec}: A  recommendation model that utilizes Gated Recurrent Units to model user interaction sequences.
    \item \textbf{SASRec}~\cite{kang2018sasrec}: A self-attention based recommendation model that employs a unidirectional Transformer to capture long-term dependencies in user behavior.
    \item \textbf{BERT4Rec}~\cite{sun2019bert4rec}: A sequential model based on the bidirectional Transformer architecture. It utilizes the Cloze objective to learn bidirectional context representations.
\end{itemize}

\paragraph{LLM-based Recommendation}
These methods directly leverage LLMs for recommendation.
\begin{itemize}[leftmargin=*]
    \item \textbf{LLM (Base)}: The backbone model used with the hierarchical agentic reasoning. It predicts the next category directly based on the prompt without any parameter updates, serving as a baseline to assess the raw capability of the model.
    \item \textbf{LLM+SFT}: The backbone model fine-tuned via standard Supervised Fine-Tuning on the training data. It treats recommendation as a standard next-token prediction task without the  agentic reasoning or reinforcement learning components.
\end{itemize}

\paragraph{LLM-Enhanced Recommendation}These methods leverage LLMs to generate auxiliary knowledge signals, such as user profiles, item descriptions, or reasoning chains. These semantic signals are then encoded and injected into conventional sequential backbones to alleviate data sparsity and improve representation robustness.
\begin{itemize}[leftmargin=*]
    \item \textbf{KAR}~\cite{xi2023KAR}: Knowledge Augmented Reasoning for recommendation. This method represents the LLM-as-Enhancer paradigm, which augments a BERT4Rec backbone with two distinct types of LLM-derived knowledge: (1) Factual Representation, obtained by encoding the semantic embeddings of explicit user living needs, and (2) Reasoning Representation, which captures the LLM's dynamic reasoning process as it infers user living need based on the current spatiotemporal context.
\end{itemize}  

\subsection{Details of dataset}
~\label{details of dataset}

The detailed statistics of both datasets are summarized in Table ~\ref{tab:dataset_stat}.
\begin{table}[h]
\centering
\caption{Statistics of the Shanghai and Beijing Datasets.}
\label{tab:dataset_stat}
\setlength{\tabcolsep}{6pt} 
\begin{tabular}{lrr}
\toprule
\textbf{Metric} & \textbf{Shanghai} & \textbf{Beijing} \\
\midrule
Total Users & 10,302 & 11,889 \\
Total Categories & 422 & 408 \\
Total Interactions & 263,437 & 207,492 \\
Avg. Sequence Length & 25.57 & 17.45 \\
Sparsity & 93.94\% & 95.73\% \\
\bottomrule
\end{tabular}
\end{table}

\subsection{Performance on Cold-Start Users(RQ3)}

To evaluate the effectiveness of our framework in cold-start scenarios, we construct a strictly defined cohort of 1,943 new users, each with only two historical interactions. We compare HiAgentRec against a range of baselines.
As shown in Table~\ref{tab:cold_start}, HiAgentRec achieves the best performance across all metrics, with an HR@5 of 0.4260 and NDCG@5 of 0.4288. 

\begin{table}[t]
\centering
\caption{Performance comparison on Cold-Start users ($N=1,943$ users with sequence length $= 2$). HiAgentRec demonstrates superior capability in handling unseen users. The best results are \textbf{bold}, and the second-best are \underline{underlined}.}
\vspace{-0.3cm}
\label{tab:cold_start}
\resizebox{\columnwidth}{!}{
\begin{tabular}{l ccccc}
\toprule
\textbf{Model} & HR@1 & HR@3 & HR@5 & NDCG@3 & NDCG@5 \\
\midrule
GRU4Rec & 0.1450 & 0.2050 & 0.2501 & 0.1820 & 0.1591 \\
SASRec & 0.2676 & 0.2910 & 0.3139 & 0.2840 & 0.2933 \\
BERT4Rec & 0.2676 & 0.2820 & 0.2908 & 0.2750 & 0.3264 \\

LLM (Base) & 0.0420 & 0.0510 & 0.0610 & 0.0450 & 0.0502 \\
LLM+SFT & 0.0480 & 0.0650 & 0.0861 & 0.0520 & 0.0640 \\
KAR & \underline{0.2950} & \underline{0.3180} & \underline{0.3310} & \underline{0.3080} & \underline{0.3450} \\
\midrule
\textbf{HiAgentRec} & \textbf{0.3759} & \textbf{0.4080} & \textbf{0.4260} & \textbf{0.3950} & \textbf{0.4288} \\
\bottomrule
\end{tabular}
}
\end{table}
\subsection{Can HiAgentRec preserve general capabilities during specialization?(RQ5)}
\begin{table}[htbp]
\centering
\caption{Accuracy of models on General Benchmark.}
\vspace{-0.3cm}
\label{tab:generalization_comparison}
\resizebox{\columnwidth}{!}{
\begin{tabular}{l S[table-format=1.4] S[table-format=1.4] S[table-format=1.4]}
\toprule
Benchmark & {LLM} & {LLM+SFT} & {\textbf{HiAgentRec}} \\
\midrule
GSM8K  & 0.6808 & 0.5807 & 0.7157 \\
MMLU   & 0.6656 & 0.6616 & 0.6644 \\
IFEval & 0.4122 & 0.4104 & 0.4270 \\
MBPP   & 0.4633 & 0.4796 & 0.4776 \\
\bottomrule
\end{tabular}
}
\end{table}

We evaluate the performance of different models on general-purpose benchmarks to investigate whether our method forgets its general capabilities while specializing in a specific task. As shown in Table ~\ref{tab:generalization_comparison}, we compare the base LLM, its supervised fine-tuned variant, and our proposed method HiAgentRec across four key dimensions: GSM8K (mathematical reasoning), MMLU (factual knowledge), IFEval (instruction following), and MBPP (code generation).

The results reveal a clear contrast in generalization behavior. LLM+SFT exhibits a notable performance drop on GSM8K, suggesting that conventional supervised fine-tuning may impair complex reasoning abilities. In contrast, HiAgentRec maintains strong performance across all benchmarks. These gains may stem from the structured prompting and consistency constraints inherent in our framework, which indirectly reinforce general reasoning patterns. In summary, HiAgentRec achieves task-specific specialization while retaining general purpose capabilities.

\subsection{Theoretical Analysis(RQ6)}
\label{subsec:policy_degeneration}

Real-world recommendation environments are characterized by stochastic user behaviors and sparse rewards. In this section, we analyze the instability of GRPO when lacking hierarchical guidance. Empirically, we observe that removing the \textit{Living Need Inference} stage—forcing the model to predict semantic categories directly—leads to rapid policy collapse. We provide a theoretical derivation for this phenomenon and demonstrate how our hierarchical framework mitigates these risks via variance reduction.

Consider a policy $\pi_\theta$ optimized via GRPO. The gradient estimate for the objective function $J(\theta)$ is given by:
\begin{equation}
    \nabla J \approx \frac{1}{N} \sum_{i=1}^N \sum_{t} A_t \cdot \nabla_\theta \log \pi_\theta(a_t|s_t)
\end{equation}
In a sparse-reward setting without intermediate semantic anchors, the advantage term $A_t$ exhibits extreme variance. 
When the environment yields a stochastic positive signal for a noisy action sequence $a_{\text{noise}}$, the gradient estimator lacks the inductive bias to distinguish valid reasoning from serendipitous noise. Consequently, the optimizer indiscriminately amplifies the probability of these shortcut patterns.
As the policy overfits to these high-variance signals, the distribution converges prematurely to a deterministic state, characterized by vanishing entropy:
\begin{equation}
    H(\pi_\theta) \to 0 \quad \implies \quad \pi_\theta(a|s) \to \delta(a - a_{\text{degenerate}})
\end{equation}
This mathematical collapse manifests empirically as the generation of repetitive or nonsensical tokens ("gibberish"), as the model gets trapped in poor local optima.

HiAgentRec prevents this degeneration by decomposing the direct mapping $P(b|s)$ into a conditional probability chain: $P(b|c, i, s) \cdot P(c|i, s) \cdot P(i|s)$.
This hierarchical structure leverages the \textit{Law of Total Variance} to constrain the optimization space. The variance of the target prediction $B$ given state $S$ can be decomposed as:
\begin{equation}
    \text{Var}(B|S) = \underbrace{\mathbb{E}[\text{Var}(B|C,I,S)]}_{\text{Intra-cluster Variance}} + \underbrace{\text{Var}(\mathbb{E}[B|C,I,S])}_{\text{Inter-cluster Variance}}
\end{equation}
By explicitly supervising the \textit{Living Need} ($I$) and \textit{Category} ($C$) in early curriculum stages, we effectively minimize the second term (variance of the expectation). This transforms the optimization landscape: instead of searching in a high-entropy global space, the policy explores $P(b|c,i,s)$ within a significantly constricted manifold, acting as a semantic regularizer that prevents collapse into degenerate regions.

\subsection{Semantic Model Comparison(RQ7)}
\begin{figure}[htbp]
\centering
\includegraphics[width=1.0\linewidth]{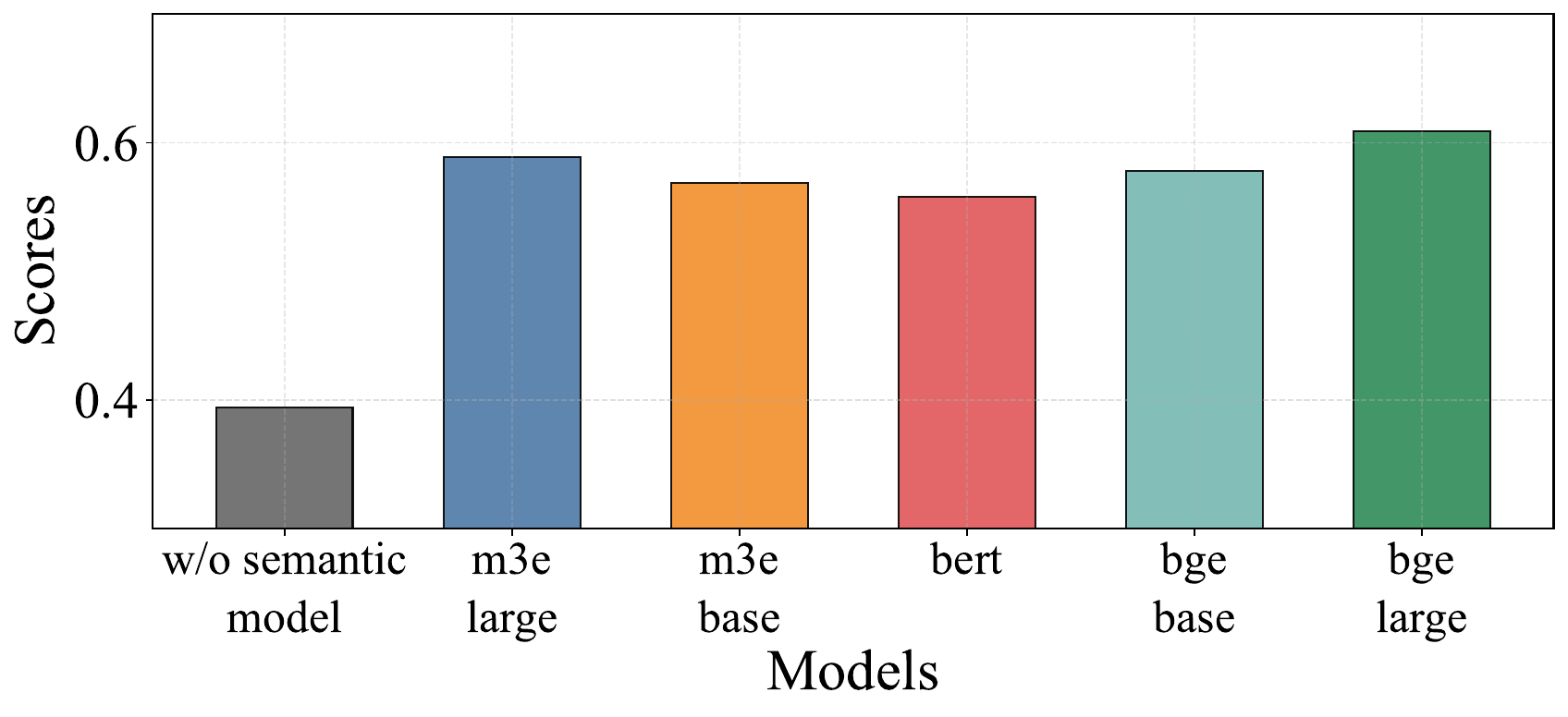}
\caption{Comparison of Category Match Accuracy(Hit Rate@1) Across Different Semantic Models}
\label{fig:semantic_cmpare}
\end{figure}

As shown in Figure ~\ref{fig:semantic_cmpare}, adopting an approach without a semantic model yields low classification accuracy (approximately 0.39), where the raw output of LLM is directly matched against predefined business categories (e.g., Hong Kong and Taiwanese Desserts) via exact string matching. This limitation arises because LLMs tend to generate natural, descriptive, and diverse phrasings (e.g., a dessert shop specializing in Taiwanese taro balls and mango pomelo sago), whereas the ground-truth category system is highly structured and fine-grained (e.g., Hong Kong and Taiwanese Desserts, Sichuan Cuisine, Japanese Food). When the output does not explicitly contain the canonical category name, exact matching fails, even if the semantic meaning is fully aligned. Moreover, local service taxonomies often exhibit synonymy, near-synonymy, or hierarchical nesting 
(e.g., Desserts vs. Hong Kong and Taiwanese Desserts), further undermining the reliability of literal matching. In contrast, semantic embedding models map both LLM outputs and category labels into a shared dense vector space, enabling robust soft matching based on semantic similarity. Notably, different embedding models exhibit varying degrees of effectiveness. 

\subsection{Agentic Reasoning Prompt Specification}
\label{sec:prompt_spec}

The complete prompt structure is illustrated in the box below:
\newpage
\begin{tcolorbox}[
  title={Agentic Reasoning Pipeline},
  enhanced,
  breakable,
  colframe=cyan!40!blue!30!white,  
  colbacktitle=cyan!10!blue!10!white, 
  coltitle=black,                  
  colback=white,                   
  fonttitle=\normalsize,       
  fontupper=\small,
  boxrule=1pt,                     
  arc=4pt,                         
  boxsep=6pt,
  left=5pt, right=5pt, bottom=5pt, top=5pt,
  shadow={0mm}{0mm}{0mm}{white}    
]

\textbf{System Role:} \\
You are an autonomous Recommendation Agent. Your objective is to formulate precise recommendations by orchestrating specific function tools to analyze user data, map semantic categories, and rank potential behaviors based on the current spatiotemporal context.

\vspace{0.4em}

\begin{tcolorbox}[title={Step 1: Living need Inference}, substep_clean]
\textbf{Agent Action:} \\
Query the UserProfile MCP to retrieve long-term preferences. Combine with current Context (Time/Location) to filter candidate living needs.

\textbf{Reasoning Requirement:} Explain which profile feature triggered the selection of the living need from the candidate list.

\textbf{Output format:}
{\small
\begin{verbatim}
<intent>
{"predicted_intent": "Intent Name",
 "reasoning_summary": "Brief Reasoning"}
</intent>
\end{verbatim}
}
\end{tcolorbox}

\begin{tcolorbox}[title={Step 2: Category Mapping}, substep_clean]
\textbf{Agent Action:} \\
Call the Intent Parser to retrieve the predicted living need from Step 1.

\textbf{Reasoning Requirement:} Justify the category choice by linking the living need to specific domain availability. 
Semantic domains for categories:
\vspace{-0.3em}
\begin{itemize}[leftmargin=*]
    \item Food \& Beverage (e.g., Chinese/Western cuisine...)
    \item Accommodation (e.g., luxury, budget hotels)
    \item Entertainment \& Leisure
    \item Lifestyle Services (e.g., beauty, laundry)
    \item Grocery \& Fresh Produce
\end{itemize}
\vspace{0.3em}

\textbf{Output format:}
{\small
\begin{verbatim}
<category>
{"predicted_category": "Category Name",
 "reasoning_summary": "Brief Reasoning"}
</category>
\end{verbatim}
}
\end{tcolorbox}

\begin{tcolorbox}[title={Step 3: Behavior Ranking}, substep_clean]
\textbf{Agent Action:} \\
Call the Category Parser to retrieve the predicted category from Step 2 based on the semantic matching score.

\textbf{Reasoning Requirement:} Explain why this behavior ranks highest.

\textbf{Output format:}
{\small
\begin{verbatim}
<behavior>
{"predicted_behavior": "Behavior Name",
 "reasoning_summary": "Brief Reasoning"}
</behavior>
\end{verbatim}
}
\end{tcolorbox}

\end{tcolorbox}

\end{document}